\documentclass{PoS}

\title{Pion spectrum using improved staggered fermions}

\ShortTitle{Pion spectrum using improved staggered fermions}

\author{\speaker{Taegil Bae} \\ 
Department of Physics and Astronomy, 
Seoul National University, Seoul, 151-747, South Korea \\ 
E-mail: \email{esrevinu@phya.snu.ac.kr}}

\author{Jongjeong Kim \\ 
Department of Physics and Astronomy, 
Seoul National University, Seoul, 151-747, South Korea \\ 
E-mail: \email{rvanguard@phya.snu.ac.kr}}

\author{Weonjong Lee \\ 
Frontier Physics Research Division and
Center for Theoretical Physics, \\
Department of Physics and Astronomy, Seoul National University,
Seoul, 151-747, South Korea \\ 
E-mail: \email{wlee@phya.snu.ac.kr}}

\author{Stephen R.~Sharpe \\ 
Department of Physics, University of Washington, Seattle,
WA 98195-1560, USA \\ 
E-mail: \email{sharpe@phys.washington.edu}}

\abstract{We present results for the pion multiplet spectrum
calculated using both unimproved staggered fermions and improved
HYP-smeared staggered fermions.
In the case of unimproved staggered fermions, 
we observe (consistent with previous work)
that ${\cal O}(a^2)$ taste symmetry breaking effects are 
large and comparable to the $\approx {\cal O}(p^2)$
contributions to their masses. Higher order ${\cal O}(a^2 p^2)$ effects
are also substantial enough to be seen.
For HYP-smeared staggered fermions, we find that taste breaking is
much reduced. The ${\cal O}(a^2)$ effects are observable, but are
noticeably smaller than those obtained with AsqTad-improved staggered
fermions, and much smaller than those obtained using unimproved
staggered fermions, while ${\cal O}(a^2 p^2)$ effects are suppressed
to such a level that we cannot observe them given our statistical
errors.
From this numerical study, we conclude that HYP staggered fermions
are significantly better that AsqTad fermions from the perspective of
taste symmetry breaking.
}

\FullConference{XXIV International Symposium on Lattice Field Theory\\
                 July 23-28 2006\\
                 Tucson Arizona, US}

\begin{document}
\section{Pion spectrum with staggered fermions}
\label{sec:pion}
%
%
%
Staggered fermions have four tastes,
leading to 16 tastes of flavor non-singlet pions.
These have the spin-taste structure
$(\gamma_5 \otimes \xi_F)$ with $\xi_F \in \{I, \xi_5, \xi_\mu,
\xi_{\mu 5}, \xi_{\mu\nu} = \frac{1}{2} [\xi_\mu, \xi_\nu]\}$.
These states fall into 8 irreducible
representations of the lattice timeslice group \cite{ref:golterman:1}:
$\{I\}$, $\{\xi_5\}$, $\{\xi_i\}$, $\{\xi_4\}$, $\{\xi_{i5}\}$,
$\{\xi_{45}\}$, $\{\xi_{ij}\}$, and $\{\xi_{i4}\}$.
The pion with taste $\xi_5$ is the Goldstone pion corresponding to
the axial symmetry which is exact when $m=0$ and is broken spontaneously.
The properties of the pion spectrum can be studied using
staggered chiral perturbation theory \cite{ref:wlee:1,ref:bernard:1}.
It is shown in Ref.~\cite{ref:wlee:1} that, at leading order
in a joint expansion in $p^2$ and $a^2$, 
the pion spectrum respects an SO(4) subgroup of the full SU(4) taste symmetry.
In other words, the taste symmetry breaking happens in two steps:
at ${\cal O}(p^2) \approx {\cal O}(a^2)$ the SU(4) taste symmetry
is broken down to the SO(4) taste symmetry, while at
${\cal O}(a^2 p^2)$ the SO(4) taste symmetry is broken down to
the discrete spin-taste symmetry $SW_4$~\cite{ref:wlee:1}.
As a consequence of this analysis, we expect that, to good approximation
the pions will lie in 5 irreducible representations of SO(4) taste symmetry:
$\{I\}$, $\{\xi_5\}$,
$\{\xi_\mu\}$, $\{\xi_{\mu5}\}$, $\{\xi_{\mu\nu}\}$.
%

The splittings between the pion multiplets are a non-perturbative
measure of taste-symmetry breaking, and can be used to measure
the efficacy of different improvement schemes.
The more effective the improvement, the smaller the expected splitting.
In this paper, we present results for the  pion spectrum calculated using
different choices of staggered fermions---unimproved, HYP and AsqTad---and
compare their splitting patterns in order to determine which
improvement scheme is better or more efficient in reducing the taste
symmetry breaking.
%

In a recent study of $B_K$ using staggered chiral perturbation
theory \cite{ref:sharpe:1}, it was found that loop contributions
from non-Goldstone pions (with $\xi_F \ne \xi_5$) was much larger
than that from the Goldstone pion.
One motivation for the present study is to determine all the pion masses
so that they can be used as inputs into the chiral perturbation theory
fit for $B_K$.

\section{Cubic symmetry of the source}
\label{sec:cubic}
In order to select a specific pion taste and to exclude all others,
it is necessary to choose sources and/or sinks to lie in specific
representations of the timeslice group.
Here we discuss how this is done for the sources. We consider
only sources with vanishing physical three-momentum.
We adopt two different methods,
one the ``cubic U(1) source'' and the other the ``cubic wall-source''.
We first fix gauge configurations to Coulomb gauge.
Then propagators are obtained, as usual, by solving the following
Dirac equation with $h$ set to a specific source type.
\begin{eqnarray}
& & (D + m) \chi(x) = h(y) \\
& & \chi(x,a) = 
\sum_{y,b} G(x,a;y,b) h(y,b) 
\end{eqnarray}
where $a,b$ are color indices and $G(x;y)$ is a quark propagator.
For the cubic U(1) source at time slice $t$ we
choose $h(y)$ as follows:
\begin{eqnarray}
& & h(y,b) = \delta_{y_4,t}
\delta^3_{\vec{y}, 2\vec{n}+\vec{A}} \xi(\vec{n},b)\ , 
\mbox{ and } \xi(\vec{n},b) \in U(1)
\\ & & 
\lim_{N \rightarrow \infty} \frac{1}{N}
\sum_{\xi} \xi(\vec{n},c)  \xi^{*}(\vec{n}',c')
= \delta_{\vec{n},\vec{n}'} \delta_{c,c'}
\end{eqnarray}
where $\vec{A}$ is a cubic vector $\in \{(0,0,0), (1,0,0), \cdots,
(1,1,1)\}$, $N$ is the number of random vectors and $\vec{n} \in Z^3$.
Similarly, in order to set up the cubic wall source
at time slice $t$, we define $h(y,b)$ as follows:
\begin{eqnarray}
& & h(y,b) = \delta_{y_4,t}
\delta^3_{\vec{y}, 2\vec{n}+\vec{A}} \xi(b) \ ,
\mbox{ and } \xi(b) \in U(1)
\\ & & 
\lim_{N \rightarrow \infty} \frac{1}{N}
\sum_{\xi} \xi(c)  \xi^{*}(c') = \delta_{c,c'}
\end{eqnarray}
In other words, the random vectors are the same on all spatial
hypercubes. Note that there are eight sources of each type,
depending on the choice of $\vec A$. By combining these we
can project 
onto particular representations of the timeslice group.
We do not know, however, which type of source is more efficient; 
this can only be determined through a numerical study.

\begin{table}[t!]
\begin{center}
\begin{tabular}{ c | c }
\hline
parameter & value \\
\hline
$\beta$  & 6.0 (quenched QCD, Wilson plaquette action)  \\
$1/a$ & 1.95 GeV \\
geometry & $16^3 \times 64$  \\
\# of confs & 218  \\
gauge fixing & Coulomb  \\
bare quark mass & 0.005, 0.01, 0.015, 0.02, 0.025, 0.03  \\
$Z_m$ & $\approx 2.5$  \\
\hline
\end{tabular}
\end{center}
\caption{Simulation parameters for unimproved staggered fermions}
\label{tab:par:unimp}
\end{table}

\section{Operator Transcription}
\label{sec:op}
We use two different methods to construct bilinear operators: 
the ``Kluberg-Stern method'' \cite{ref:klu:1} and 
the ``Golterman method'' \cite{ref:golterman:1}.
In the Kluberg-Stern method, a bilinear operator is expressed as
\begin{eqnarray}
& & O_{S,F} = \sum_{A,B} \bar{\chi}(A) 
\overline{(\gamma_S \otimes \xi_F)}_{A,B}  \chi(B) \\
& & \overline{(\gamma_S \otimes \xi_F)}_{A,B}
= \frac{1}{4} {\rm Tr} ( \gamma_A^\dagger \gamma_S \gamma_B \gamma_F^\dagger )
\end{eqnarray}
where $A,B$ is a hypercubic vector $\in \{(0,0,0,0), (1,0,0,0),
\cdots, (1,1,1,1)\}$, $\gamma_S$ ($\xi_F$) represents the spin
(taste). The operator is made gauge invariant
by fixing each timeslice to Coulomb gauge, so
that spatial links are not required, and inserting appropriate
time-directed links if $S_4\ne F_4$.
In the Golterman method, the bilinears are
\begin{eqnarray}
& & O_{S,F} =  \eta_{S,F}(x) \bar{\chi}(x) M_{S,F} \chi(x) \\
& & M_{S,F} \chi(x) = \prod_{\mu=1,2,3}
\Big[ (1 - |S_\mu - F_\mu|) + |S_\mu - F_\mu| D_\mu \Big] \chi(x) \\
& & D_\mu \chi(x) = \frac{1}{2} \Big[ \chi(x + \hat{\mu})
+ \chi(x - \hat{\mu}) \Big]
\end{eqnarray}
where $\eta_{S,F}(x)$ is a phase factor given in
Ref.~\cite{ref:golterman:1}, and gauge invariance is maintained
as for the Kluberg-Stern operators.

The Kluberg-Stern method gives bilinears that are only
approximate representations of the timeslice group whereas the
Golterman method sorts out the bilinears according to true irreducible
representations \cite{ref:golterman:1}.
We apply both methods to our numerical study and the results will be
compared in the next section.

\section{Numerical Study with unimproved staggered fermions}
\label{sec:num:unimp}
\begin{figure}[t!]
\includegraphics[width=0.5\textwidth]{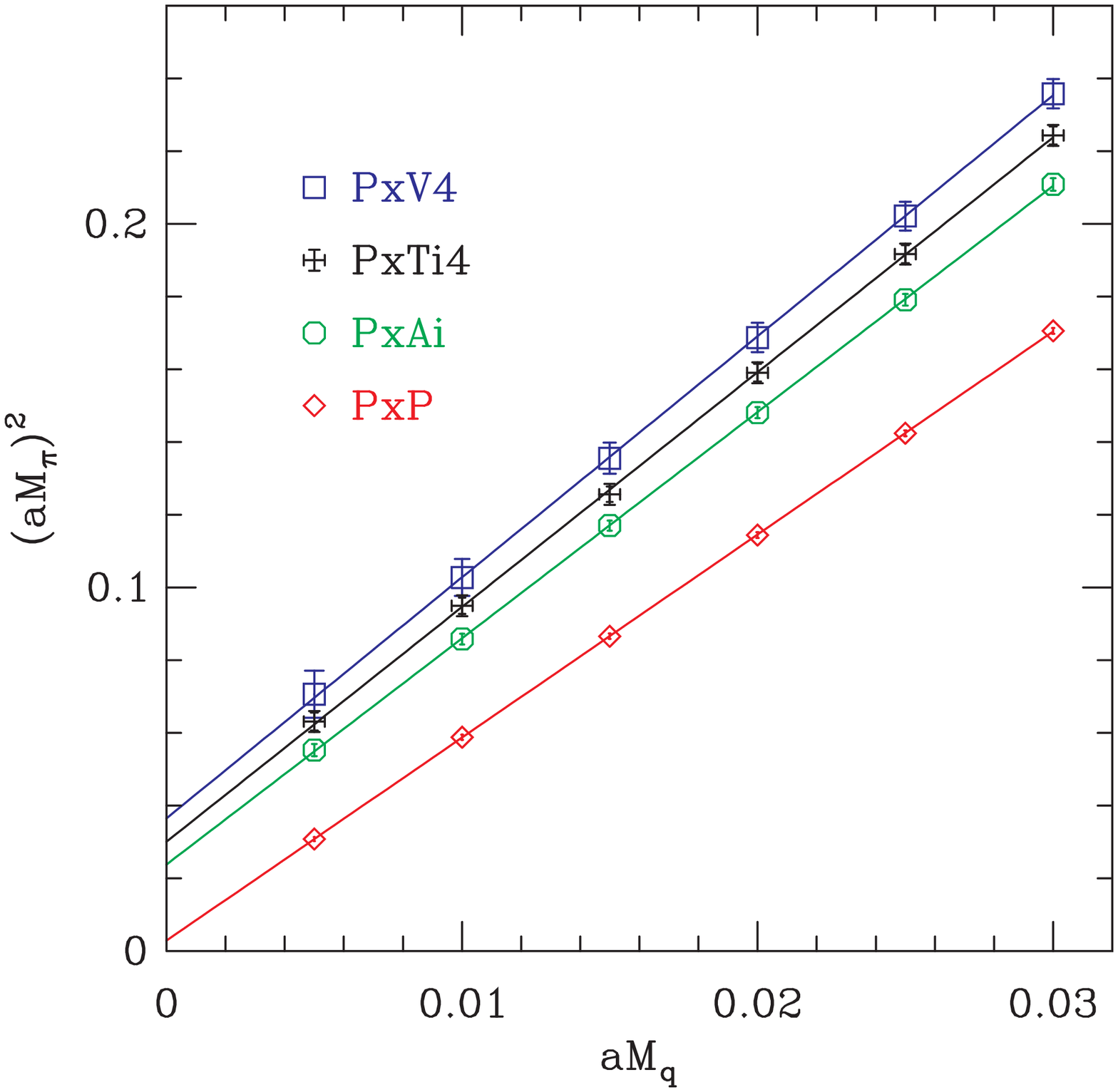}
\includegraphics[width=0.5\textwidth]{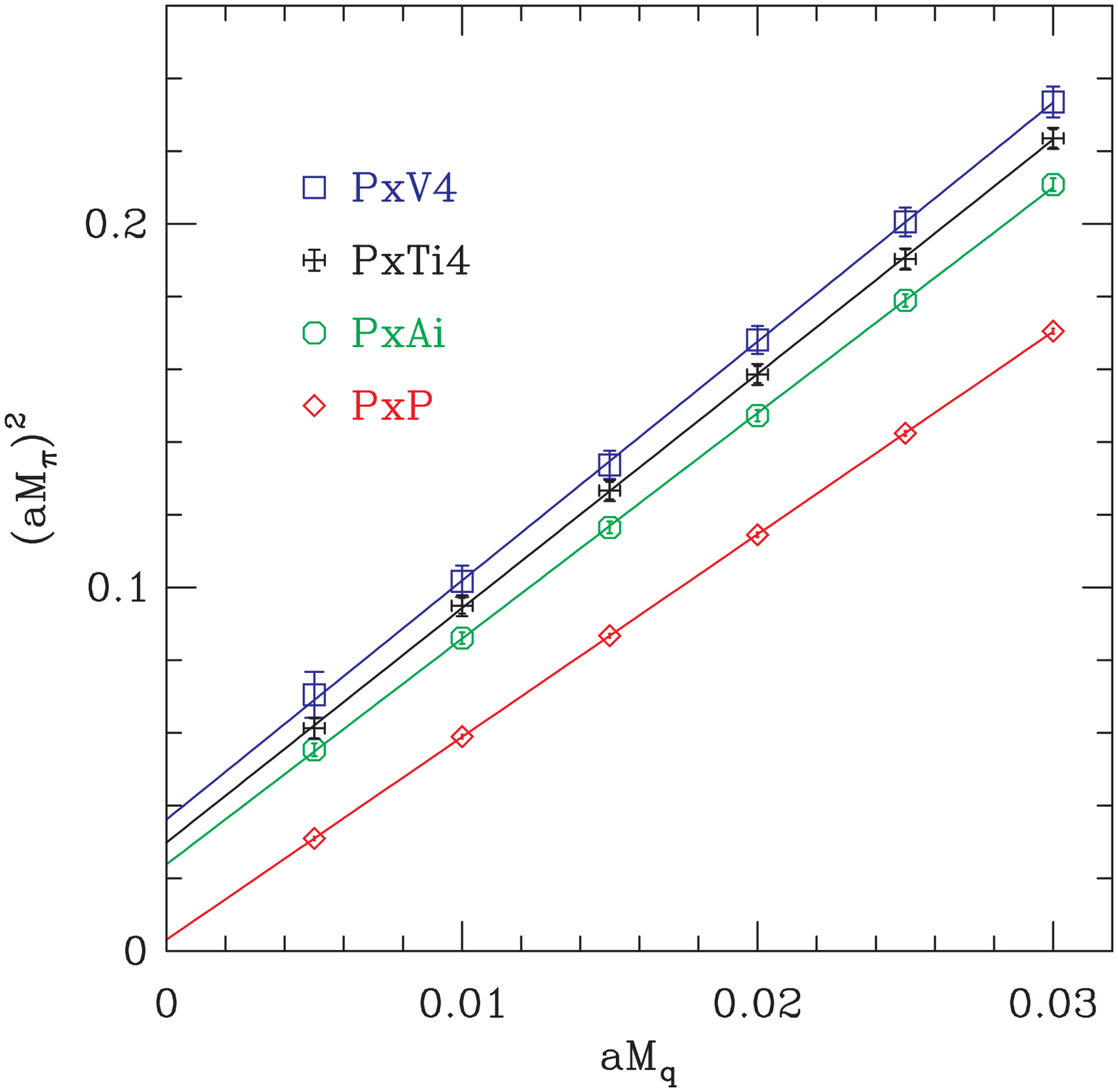}
\caption{$(a m_\pi)^2$ vs.~$am_q$ for unimproved staggered fermions: 
  (left) using the Kluberg-Stern method;
  (right) using the Golterman method; 
  In both calculations, the sources are set to cubic U(1) source.}
\label{fig:mpisq:unimp}
\end{figure}
The parameters for the numerical study using unimproved staggered 
fermions are summarized in Table~\ref{tab:par:unimp}.
In Fig.~\ref{fig:mpisq:unimp} we show $(am_\pi)^2$ as a function of
quark masses using bilinear operators with spin-taste $\gamma_5\otimes
\xi_4$, $\gamma_5\otimes \xi_{i4}$, $\gamma_5\otimes \xi_{i5}$, and
$\gamma_5\otimes \xi_5$.
We find essentially no difference between the
Kluberg-Stern method (left panel) and the
Golterman method (right panel).
The Kluberg-Stern operator includes the Golterman operator as the
leading term, as well as operators with derivatives having different
spin-tastes. The latter couple to different states which should be
projected against by the sources. Nevertheless, one might expect
the Kluberg-Stern operators to have a more noisy signal.
In fact, we see that the signals with the two methods are essentially
identical.

In Fig.~\ref{fig:mpisq:unimp}, we also observe that the splitting
between the pion multiplets are comparable to the light pion masses,
implying that ${\cal O}(a^2) \approx {\cal O}(p^2)$.
In addition, we observe that the slopes are different for
various tastes, implying that ${\cal O}(a^2 p^2)$ terms are
significant.

\section{Numerical Study with HYP staggered fermions}
\label{sec:num:hyp}
Using the same set of gauge configurations as in
Sec.~\ref{sec:num:unimp}, we study the pion spectrum with HYP-smeared
staggered fermions \cite{ref:hyp:1}.
The parameters for the HYP fat links are set to the HYP (II) condition
in Ref.~\cite{ref:wlee:2}.
In the numerical study, we use quark masses of 0.01, 0.02, 0.03, 0.04,
0.05.
Note that $Z_m \approx 1$ for HYP staggered fermions whereas
$Z_m \approx 2.5$ for unimproved staggered fermions.
Hence, the physical quark masses in this study
are comparable to those for the unimproved
staggered fermions considered above.
\begin{figure}[t!]
\includegraphics[width=0.5\textwidth]{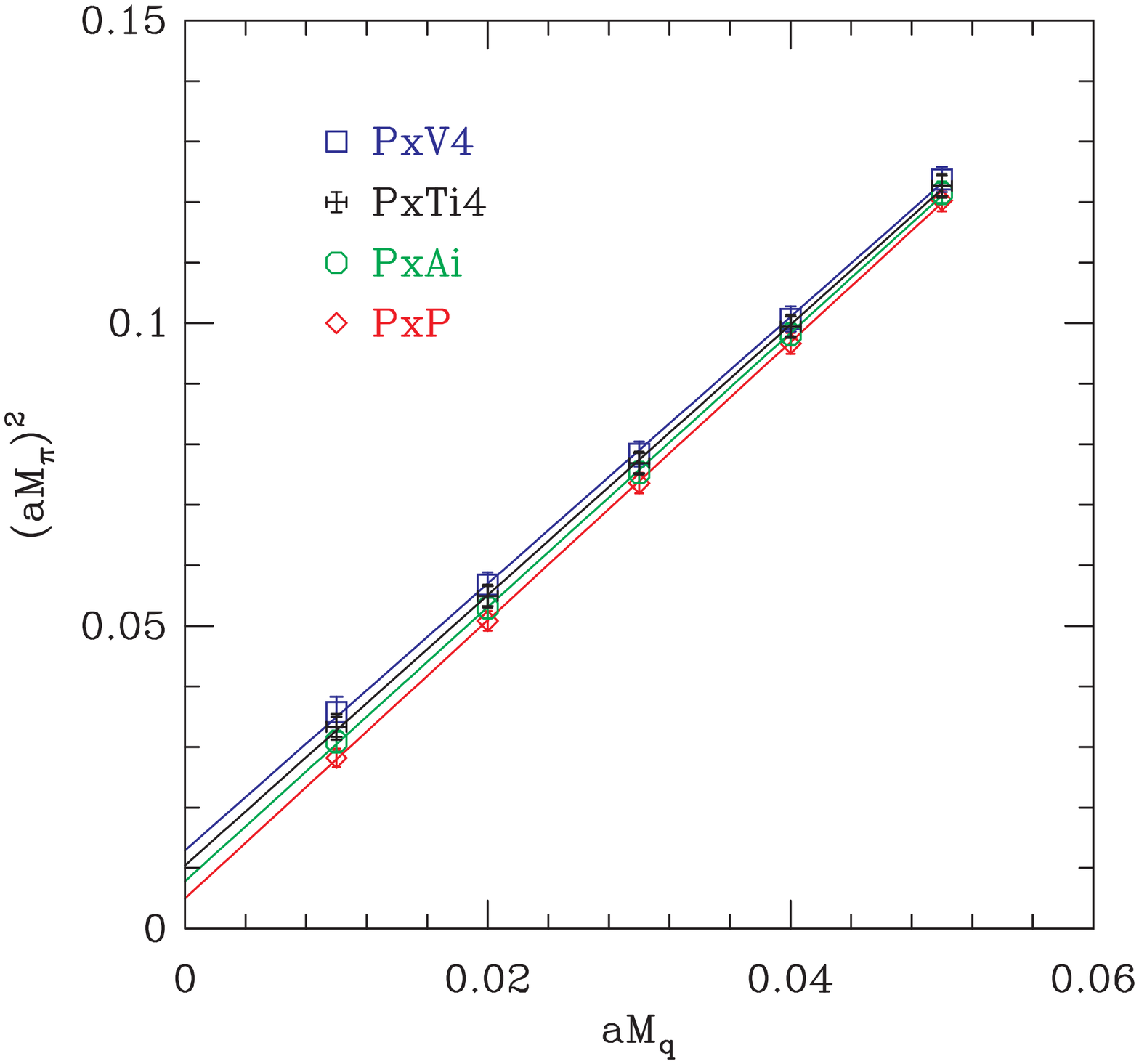}
\includegraphics[width=0.5\textwidth]{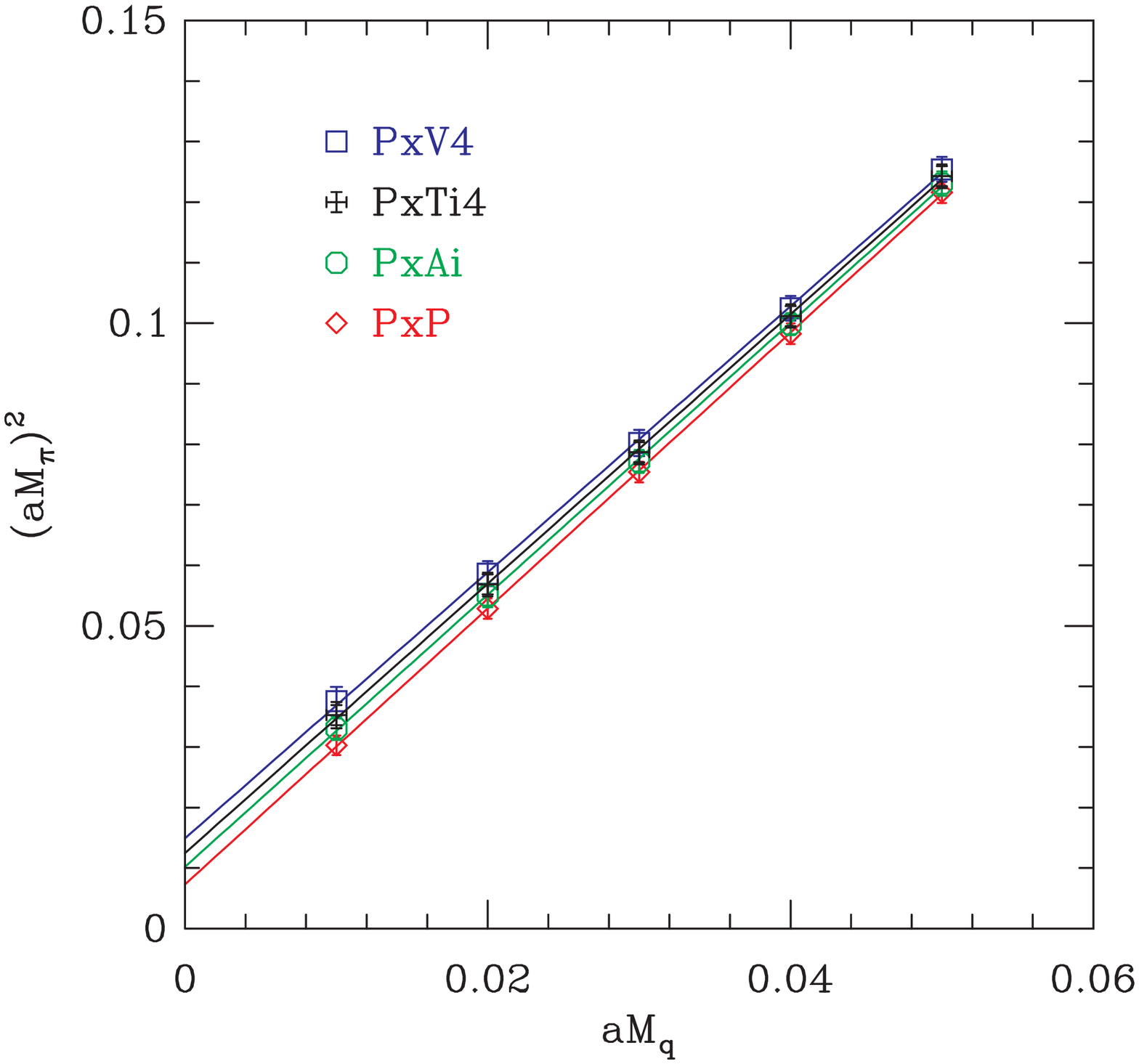}
\caption{$(a m_\pi)^2$ vs.~$am_q$ for HYP staggered fermions: 
  (left) using the cubic U(1) source;
  (right) using the cubic wall source; 
  In both calculations, the operators are constructed using
  the Golterman method.}
\label{fig:mpisq:hyp:src}
\end{figure}
In Fig.~\ref{fig:mpisq:hyp:src}, we show $(am_\pi)^2$ as a function
of quark mass and compare the results of the cubic U(1) source (left)
with those of the cubic wall source (right). 
This comparison indicates there is essentially no difference
between the two sources in practice.
In Fig.~\ref{fig:mpisq:hyp:src}, we observe that the splittings
between the pions are significantly suppressed, down to
the $2\sigma$ level in our data set, implying that ${\cal
O}(a^2) \ll {\cal O}(p^2)$.
In addition, the slopes for different tastes are
equal within statistical uncertainty, showing that improvement
reduces the ${\cal O}(a^2 p^2)$ effects as well.

\begin{figure}[t!]
\includegraphics[width=0.5\textwidth]{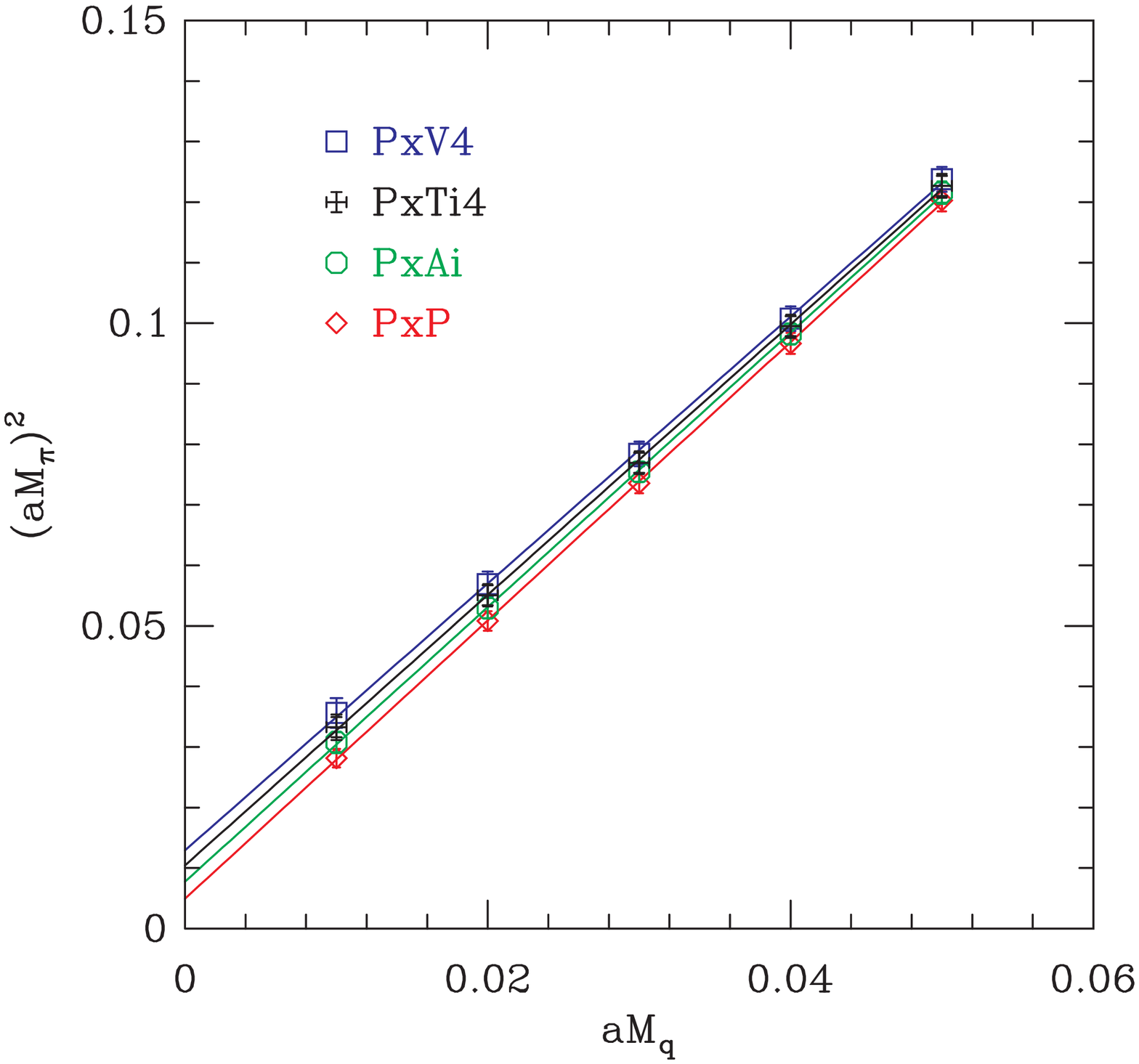}
\includegraphics[width=0.5\textwidth]{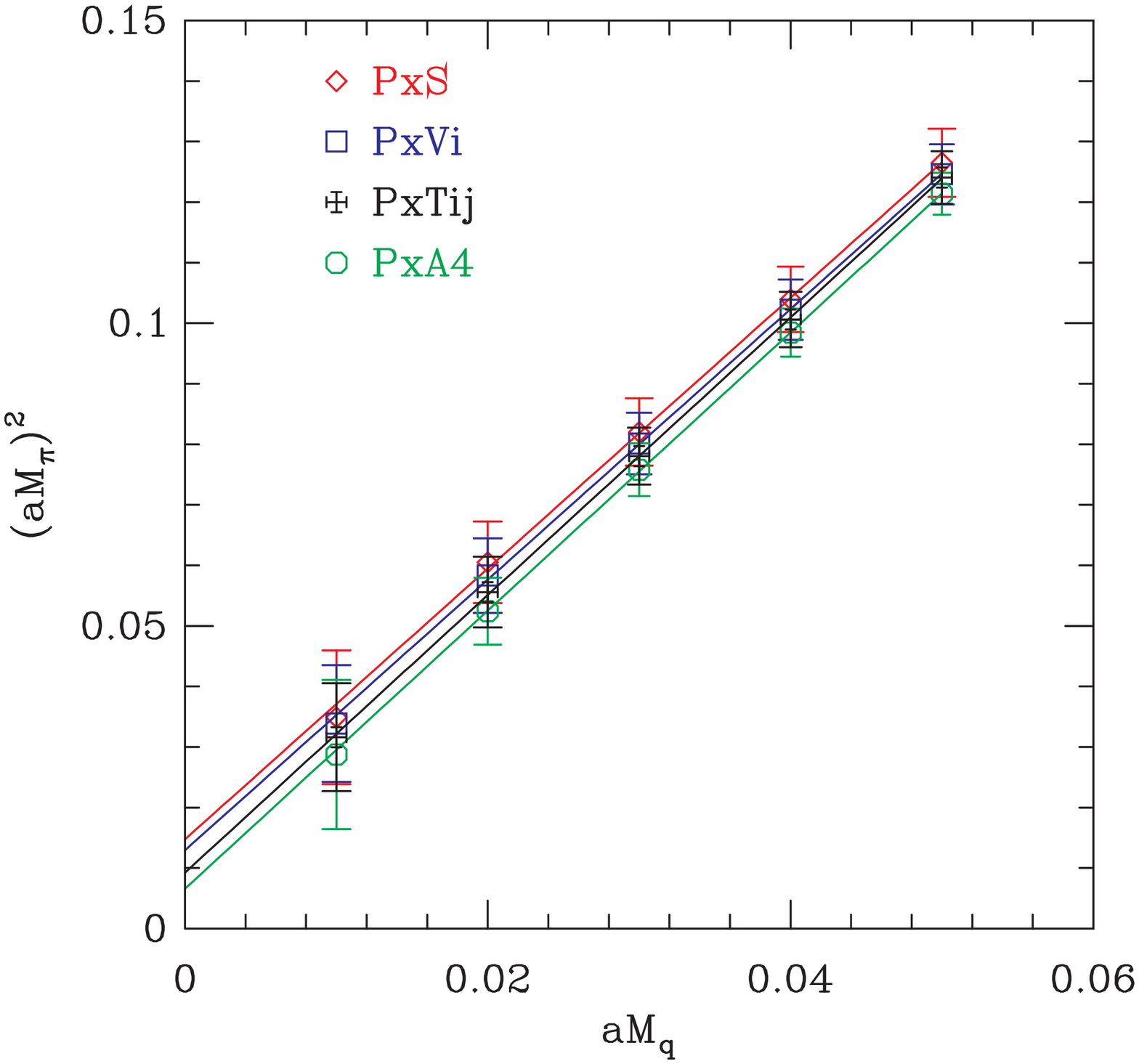}
\caption{$(a m_\pi)^2$ vs.~$am_q$: 
  (left) for the operators local in time;
  (right) for the operators non-local in time; 
  In both cases, we use the Kluberg-Stern method to construct the
  operators and we also choose cubic U(1) sources. }
\label{fig:mpisq:hyp:time}
\end{figure}
We can divide the pion bilinear operators into two categories: one is
local in time ($\gamma_5 \otimes \xi_5$, $\gamma_5 \otimes \xi_4$,
$\gamma_5 \otimes \xi_{i5}$, and $\gamma_5 \otimes \xi_{i4}$) and the
other is non-local in time ($\gamma_5 \otimes 1$, $\gamma_5
\otimes \xi_i$, $\gamma_5 \otimes \xi_{45}$, and $\gamma_5 \otimes
\xi_{ij}$).
In the case of sink operators local in time, we can construct the
source operators with the same spin and taste structure.
However, in the case of sink operators non-local in time, we cannot impose
the same spin and taste on the source operators, because the source 
is local in time by construction.
Since we want to select a pion with a specific taste, we do not have any 
other choice for the taste but we do have the freedom to choose the spin:
we may choose either pseudoscalar ($\gamma_5$) or axial vector
($\gamma_{45}$).
Hence, in the case of sink operators non-local in time, we choose
axial current as a source operator, which is local in time.
For example, if the sink operator is $\gamma_5 \otimes \xi_i$
(non-local in time), we choose $\gamma_{45} \otimes \xi_i$ (local in
time) as the source operator.
%
%
In Fig.~\ref{fig:mpisq:hyp:time}, we show $(a m_\pi)^2$ as a function
of quark mass for operators local in time (left) and for operators
non-local in time (right).
We see that the signals for the operators non-local in time are noisier than 
those for the local operators.

We also observe that the pattern of pion multiplet spectrum is in the
following order, which is the same for unimproved staggered fermions
and AsqTad staggered fermions:
\begin{equation}
m_\pi(\xi_5) < m_\pi(\xi_{\mu5}) < m_\pi(\xi_{\mu\nu})
< m_\pi(\xi_\mu) < m_\pi(I)
\end{equation}

\section{Comparison of AsqTad and HYP staggered fermions}
\label{sec:num:cmp}
\begin{figure}[t!]
\includegraphics[width=0.5\textwidth]{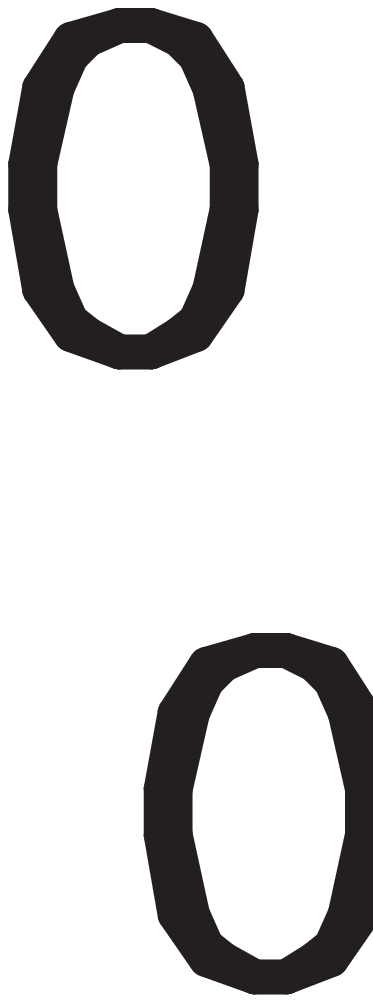}
\includegraphics[width=0.5\textwidth]{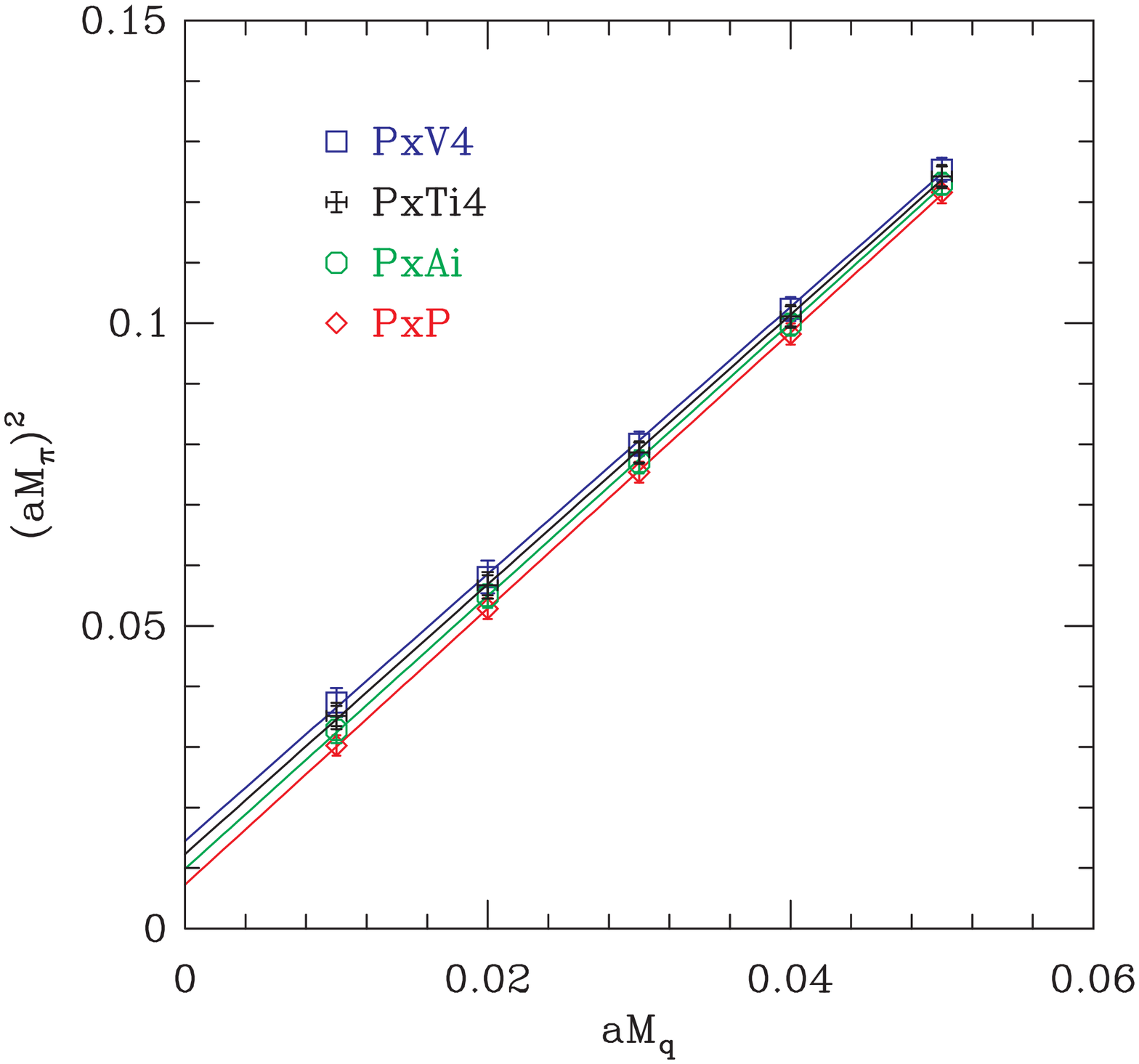}
\caption{$(a m_\pi)^2$ vs.~$am_q$:
  (left) for AsqTad staggered fermions;
  (right) for HYP staggered fermions (the Kluberg-Stern method, cubic wall 
  source) }
\label{fig:mpisq:asqtad:hyp}
\end{figure}

In Fig.~\ref{fig:mpisq:asqtad:hyp}, we compare the results of AsqTad
staggered fermions (left) with those of HYP staggered fermions (right).
The AsqTad staggered fermion data comes originally from
Ref.~\cite{ref:milc:1}.  
It has $a=0.125$ fm (MILC coarse lattice) whereas
the results for HYP-smeared staggered fermions are on
a finer lattice with $a \approx 0.1$ fm.
Nevertheless, since the gauge part of the AsqTad action is
Symanzik improved, while our calculation uses the unimproved Wilson
gauge action, 
we expect that the scaling violations from the background gauge
configurations to be similar.
The physical quark masses are also similar.

In the case of AsqTad staggered fermions, we observe that the 
taste-symmetry breaking is of the same order as the light pion masses.
In other words, ${\cal O}(a^2) \approx {\cal O}(p^2)$ for the AsqTad
action.
In addition, we notice that the slopes for different tastes are
very close, implying that this manifestation of ${\cal O}(a^2 p^2)$
terms is very small.
By contrast, with HYP-smeared staggered fermions we see that
both ${\cal O}(a^2)$ and ${\cal O}(a^2 p^2)$ effects are very small.
We conclude that HYP staggered
fermions are significantly better than AsqTad staggered fermions from the
standpoint of taste symmetry breaking.

\section{Acknowledgment}
\label{sec:ack}
W.~Lee acknowledges with gratitude that the research at Seoul National
University is supported by the KOSEF grant (R01-2003-000-10229-0), by
the KOSEF grant of international cooperative research program, by the
BK21 program, and by the US DOE SciDAC-2 program.
The work of S.~Sharpe is supported in part by the US DOE grant
no. DE-FG02-96ER40956, and by the US DOE SciDAC-2 program.


%
%
%


\begin{thebibliography}{99}
%
\bibitem{ref:golterman:1} Maarten Golterman,
\emph{Staggered Mesons}, 
\emph{Nucl.~Phys.} B{\bf 273} (1986) 663-676.
%
\bibitem{ref:wlee:1} Weonjong Lee and Stephen Sharpe, 
\emph{Partial Flavor Symmetry Restoration for Chiral Staggered Fermions}, 
\emph{Phys.~Rev.} D{\bf 60} (1999) 094503,
[{\tt hep-lat/9905023}].
%
\bibitem{ref:bernard:1} C.~Aubin and C.~Bernard, 
\emph{Pion and kaon masses in staggered chiral perturbation theory}, 
\emph{Phys.~Rev.} D{\bf 68} (2003) 034014,
[{\tt hep-lat/0304014}].
%
\bibitem{ref:sharpe:1}  Ruth S.~Van de Water, Stephen R.~Sharpe, 
\emph{$B_K$ in staggered chiral perturbation theory}, 
\emph{Phys.~Rev.} D{\bf 73} (2006) 014003,
[{\tt hep-lat/0507012}].
%
\bibitem{ref:klu:1} H.~Kluberg-Stern, {\em et al.},
\emph{Flavors of Lagrangian Suskind fermions}, 
\emph{Nucl.~Phys.}  B{\bf 220} (1983) 447;
D.~Verstegen, 
\emph{Symmetry properties of fermionic bilinears $\cdots$},
\emph{Nucl.~Phys.}  B{\bf 249} (1985) 685.
%
%
\bibitem{ref:hyp:1}  A.~Hasenfratz and F.~Knechtli,, 
\emph{Flavor symmetry and the static potential with hypercubic blocking}, 
\emph{Phys.~Rev.} D{\bf 64} (2002) 034504,
[{\tt hep-lat/0103029}].
%
\bibitem{ref:wlee:2} Weonjong Lee and Stephen Sharpe, 
\emph{Perturbative matching of staggered four-fermion operators 
with hypercubic fat links}, 
\emph{Phys.~Rev.} D{\bf 68} (2003) 054510,
[{\tt hep-lat/0306016}].
%
\bibitem{ref:milc:1} C.~Aubin, {\em et al.}, 
\emph{Light pseudoscalar decay constants, quark masses, and
low energy constants from three flavor lattice QCD}, 
\emph{Phys.~Rev.} D{\bf 70} (2004) 114501,
[{\tt hep-lat/0407028}].
%
\end{thebibliography}
\end{document}